\documentclass[final,5p,times]{elsarticle}

\usepackage{amssymb}
\usepackage{amsmath}
\usepackage{xfrac}
\usepackage{enumerate}
\usepackage[hidelinks]{hyperref}
\usepackage{xurl}
\usepackage[english]{babel}
\usepackage{microtype}
\usepackage{subcaption}
\usepackage{caption}


\journal{Nuclear Instruments and Methods in Physics Research A}

\begin{document}

\begin{frontmatter}

\title{On peculiarities of the annealing process for highly transparent silica-based aerogel tiles manufactured in Novosibirsk}

\author[a,b]{A.Yu.~Barnyakov}
\author[c]{A.F.~Danilyuk}
\author[a,d]{A.A.~Kattsin\corref{cor1}}
\cortext[cor1]{Corresponding author}
\ead{A.A.Kattsin@inp.nsk.su}
\author[a,d]{E.A.~Kravchenko}
\author[c]{A.A.~Pochtar}
\author[c]{A.Yu.~Predein}

\affiliation[a]{organization={Budker Institute of Nuclear Physics},
            addressline={Acad.~Lavrentiev Prospect 11},
            city={Novosibirsk},
            postcode={630090},
            country={Russia}}
\affiliation[b]{organization={Novosibirsk State Technical University},
            addressline={Karl~Marks Prospect 20},
            city={Novosibirsk},
            postcode={630073},
            country={Russia}}
\affiliation[c]{organization={Boreskov Institute of Catalysis},
            addressline={Acad.~Lavrentiev Prospect 5},
            city={Novosibirsk},
            postcode={630090},
            country={Russia}}
\affiliation[d]{organization={Novosibirsk State University},
            addressline={St.~Pirogova 1},
            city={Novosibirsk},
            postcode={630090},
            country={Russia}}

\begin{abstract}
Since 1986, a collaboration between the Boreskov Institute of Catalysis (BIC) and the Budker Institute of Nuclear Physics (BINP) has been producing silica aerogel blocks for Cherenkov detectors. Novosibirsk-manufactured aerogel has been employed in several experiments, including KEDR and SND (BINP, Russia), DIRAC and LHCb (CERN, Switzerland), AMS-02 (ISS), and CLAS12 RICH (Jefferson Lab, USA). This work describes key advances in the production technology of large-scale aerogel radiators used in Ring-Imaging CHerenkov (RICH) detectors. Annealing is one of the key stages in the production of highly transparent aerogel in Novosibirsk. This process was studied in detail and optimized to improve the yield of aerogel tiles suitable for RICH detectors. The optical and mechanical properties of the largest silica aerogel samples produced in Novosibirsk using the new annealing procedure are presented.
\end{abstract}


\end{frontmatter}

%\linenumbers

%%=============================================================================
\section{Introduction}
\label{sec:intro}

The production of aerogel for Cherenkov counters began in Novosibirsk in 1986. Novosibirsk aerogel was used in the construction of threshold Cherenkov counters for the KEDR experiment~\cite{Barnyakov2002} (aerogel $n = 1.05$, total volume $\sim$ 1000~L), the SND experiment~\cite{Barnyakov2009} (aerogel $n = 1.13$, total volume $\sim$ 5.5~L), and the DIRAC-II experiment~\cite{Allkofer2008} (aerogel $n = 1.008$, total volume $\sim$ 9~L).
Since 2004, aerogel from Novosibirsk has been utilized in several RICH detectors, including the RICH1 detector of the LHCb experiment~\cite{Bellunato2009} (aerogel $n = 1.03$, area $\sim$ 0.5~m$^2$, tile size 200~$\times$~200~$\times$~50~mm$^3$), the RICH counter for the AMS-02 experiment~\cite{Buenerd2005} (aerogel $n = 1.05$, area $\sim$ 1~m$^2$, tile size 115~$\times$~115~mm$^2$), and the RICH detector of the CLAS12 experiment~\cite{Contalbrigo2011} (aerogel $n = 1.05$, area $\sim$ 6~m$^2$, tile size 200~$\times$~200~$\times$~30--20~mm$^3$).

For RICH detectors, in addition to optical properties, the geometrical parameters of the radiators are critically important: flatness, thickness tolerance, and dimensions. Imperfections at the radiator edges can lead to a loss of Cherenkov photons or distortions in their spatial distribution, thereby degrading particle identification performance. Therefore, it is desirable to minimize the total length of the boundaries between individual radiators and to maximize the area of monolithic elements. Consequently, the production of aerogel tiles with the largest possible lateral dimensions is essential.
The production cycle of high-transparency Novosibirsk aerogel consists of three main stages:
\begin{itemize}
	\item \textbf{Chemical synthesis} — the synthesis of the gel, with the pores filled by alcohol, during which the key material parameters are defined;
	\item \textbf{Supercritical drying} — the removal of alcohol from the pores, the stage at which the aerogel structure is formed;
    \item \textbf{Primary annealing} — a process during which maximum transparency is achieved by removing residual organic solvents from the pore structure.
\end{itemize}
In 2022--2023, a detailed study of the primary annealing process was carried out. This made it possible to optimize the procedure and significantly increase the yield of large aerogel samples suitable for use in RICH detectors. As a result, the first production of monolithic blocks measuring 230~$\times$~230~$\times$~40~mm$^3$ with a refractive index of $n = 1.05$, as well as four-layer focusing blocks with a maximum refractive index of $n_{\max} = 1.05$ and dimensions of 230~$\times$~230~$\times$~35~mm$^3$, was achieved in 2023.

%%=============================================================================
\section{Analysis of aerogel during thermal processing}

\begin{figure}[htb]
\centering
\includegraphics[width=\linewidth]{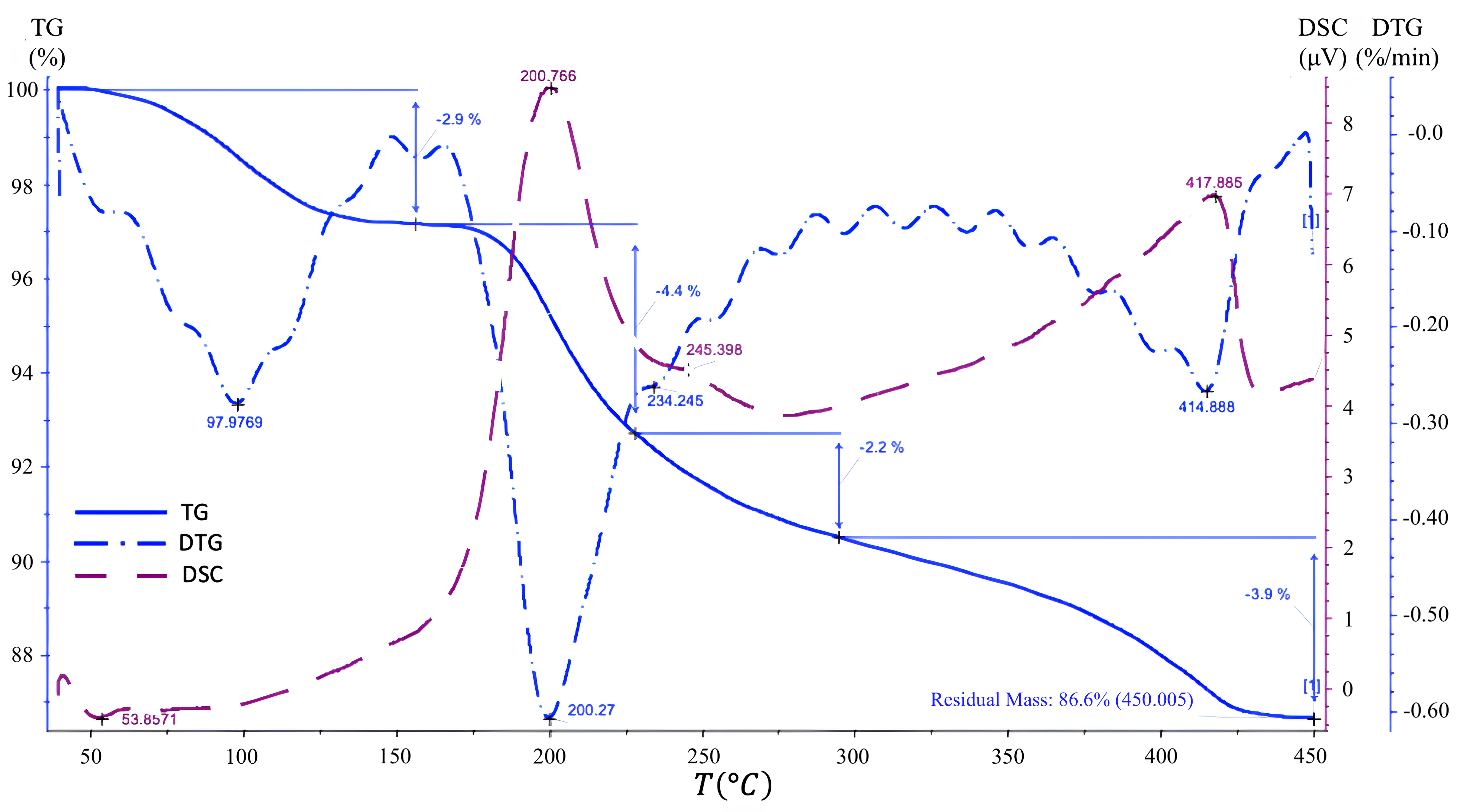}
\caption{TG (solid line), DTG (dash-dotted line) and DSC (dashed line) curves of silica aerogel combustion.}
\label{fig:fig1}
\end{figure}

Large silica aerogel tiles could occasionally fracture during the primary annealing stage, which is intended to remove alcohol groups chemically bonded to the aerogel surface, organic impurities, and residual chemicals remaining after supercritical drying. To investigate this issue, thermogravimetric analysis (TGA) and differential scanning calorimetry (DSC) were performed using a NETZSCH STA 449 C Jupiter instrument.
TGA measures the mass change of a sample during controlled heating, while DSC monitors heat flow to detect endothermic (heat absorption) and exothermic (heat release) processes such as organic decomposition or phase transitions.
Ground silica aerogel samples (10.82~mg) were heated from 22~$^\circ$C to 450~$^\circ$C at 5~$^\circ$C/min in air. The mass loss (TG), its derivative (DTG), and the heat flow (DSC) were recorded (see Fig.~\ref{fig:fig1}).

When the silica aerogel is heated, the DTG curve shows a peak at $97\,^\circ$C, which corresponds to the desorption of water and volatile components. This process is accompanied by an endothermic effect on the DSC curve.

In the temperature range $150$--$300\,^\circ$C, there is a gradual loss of mass due to burnout of residual organic impurities. This process is accompanied by a pronounced exothermic peak on the DSC curve at $200\,^\circ$C.

Further mass loss slows down, and at $417\,^\circ$C an additional exothermic peak appears on the DSC curve. The appearance of this peak is probably related to the oxidation of residual carbonaceous deposits at the final stage of heat treatment. Based on these results, the annealing procedure was modified to reduce cracking (see Section~\ref{sec:anneal}). The main hypothesis is that cracking occurs in regions where heat release is particularly rapid.

%%=============================================================================
\section{Optimization of the primary annealing protocol}
\label{sec:anneal}

The revised annealing protocol consists of slow, controlled heating steps:

\begin{enumerate}[(1)]
    \item Heat to 100~$^\circ$C at 4~$^\circ$C/h (20~h),
    \item Then to 120~$^\circ$C at 3~$^\circ$C/h (6.5~h),
    \item Then to 160~$^\circ$C at 1.5~$^\circ$C/h (26.5~h),
    \item Hold at 160~$^\circ$C for 500~min (organic burnout),
    \item Heat to 470~$^\circ$C at 5~$^\circ$C/h (62~h),
    \item Hold at 470~$^\circ$C for 200~min (final combustion),
    \item Cool to 50~$^\circ$C at 10~$^\circ$C/h (42~h).
\end{enumerate}

Total annealing time: approximately 169~h (about 7 days). For comparison, the previous annealing protocol was significantly shorter: it consisted of a constant heating rate of 5~$^\circ$C/h up to 470~$^\circ$C, followed by a 2~h hold at this temperature and cooling at 10~$^\circ$C/h. The total duration of the previous process was approximately 134~h (about 5.6 days). The new protocol enables gradual removal of organic components while preserving the structural integrity of the aerogel. Using this method, we fabricated four-layer focusing aerogel tiles (230~$\times$~230~$\times$~35~mm$^3$) for the first time. Some characteristics are given in Section~\ref{sec:multi}. We also produced large-format aerogel blocks with refractive index of 1.05 and thickness of 40~mm (see Section~\ref{sec:thick}).

%%=============================================================================
\section{The largest multilayer aerogel radiator}
\label{sec:multi}

\begin{figure}[htb]
\centering
\includegraphics[width=\linewidth]{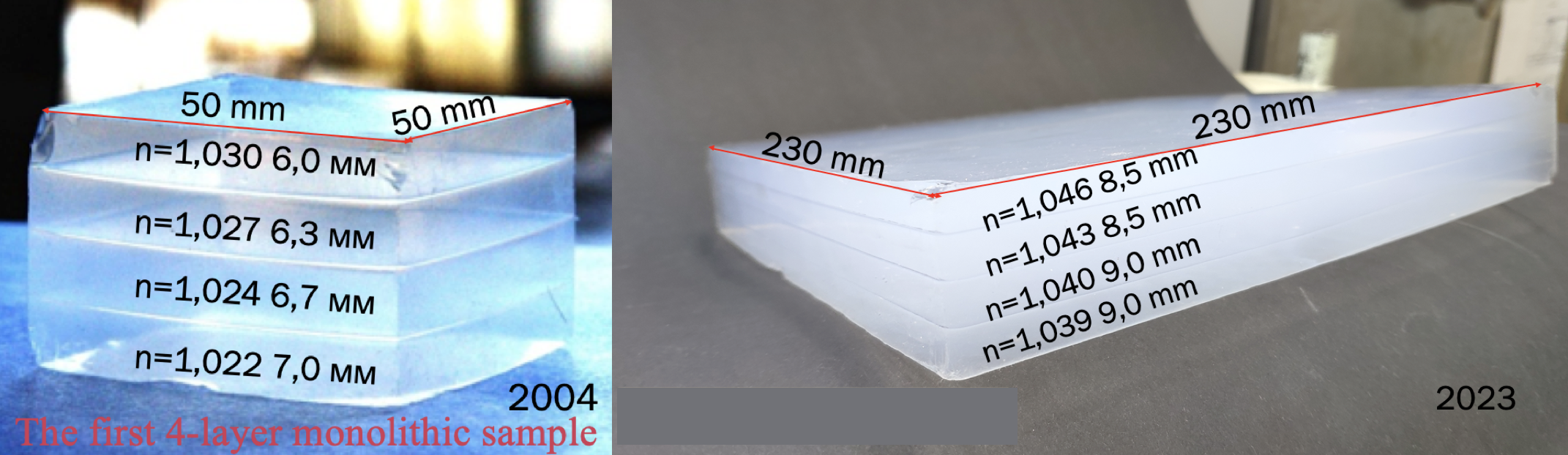}
\caption{The first four-layer sample produced in 2004 (left). The four-layer tile produced in 2023 (right).}
\label{fig:fig2}
\end{figure}

In multilayer aerogels, layer refractive indices and thicknesses are chosen so that Cherenkov rings from all layers overlap at the photon detector plane. The Novosibirsk group pioneered the production of monolithic multilayer aerogels with controlled refractive indices~\cite{Barnyakov2005}. The first multilayer monolithic aerogel block produced in 2004 is shown in Fig.~\ref{fig:fig2} (left). Building on this experience, our 2023 advancements include the successful fabrication of several four-layer focusing aerogel samples (230~$\times$~230~$\times$~35~mm$^3$), which meet the targeted specifications for refractive index and transparency (see Fig.~\ref{fig:fig2} (right)).

The refractive index of aerogels produced in Novosibirsk is related to density by the empirical relation~\cite{Danilyuk2002}
\begin{equation}
n=\sqrt{1 + 0.438\rho},
\end{equation}
where $n$ is the refractive index of the aerogel, $\rho$ is the aerogel density in g/cm$^3$.

\begin{figure}[htb]
\centering

\begin{subfigure}[b]{0.35\textwidth}
    \centering
    \includegraphics[width=\linewidth]{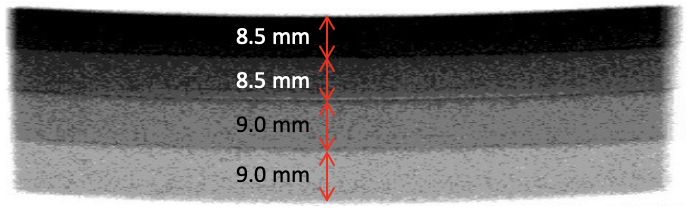}
    \caption{X-ray image of aerogel tile; darker areas indicate higher density.}
    \label{fig:xray}
\end{subfigure}
\hfill
 
\begin{subfigure}[b]{0.35\textwidth}
    \centering
    \includegraphics[width=\linewidth]{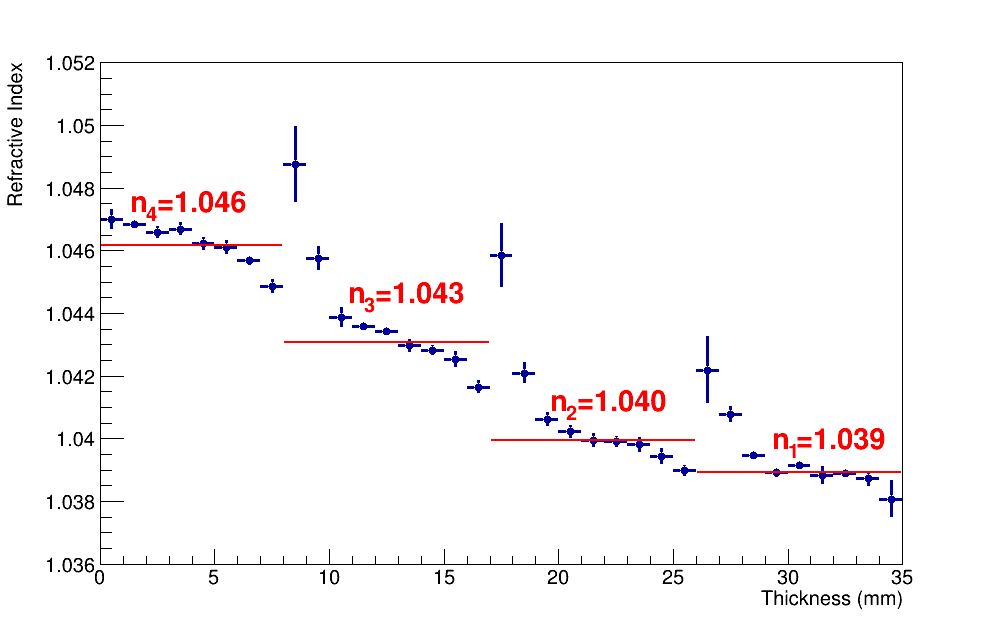}
    \caption{Profile of aerogel tile.}
    \label{fig:profile}
\end{subfigure}

\caption{X-ray image and profile of the 461f10 aerogel tile (side view).}
\label{fig:fig3}
\end{figure}

Layer indices are measured by an X-ray method~\cite{Barnyakov2006}. The X-ray image showing the thickness distribution profile of the refractive index is presented in~Fig.~\ref{fig:fig3}. Several large four-layer aerogel tiles were produced. The experimentally observed variations in these parameters allow for an assessment of the precision and tolerances of the multilayer aerogel manufacturing technology. The mean measured refractive index ($n_{i}$) values for the specific layers of the four-layer aerogels 461f10 and 461f5 are given in Table~\ref{tab:1}.

\begin{figure}[htb]
\centering
\includegraphics[width=\linewidth]{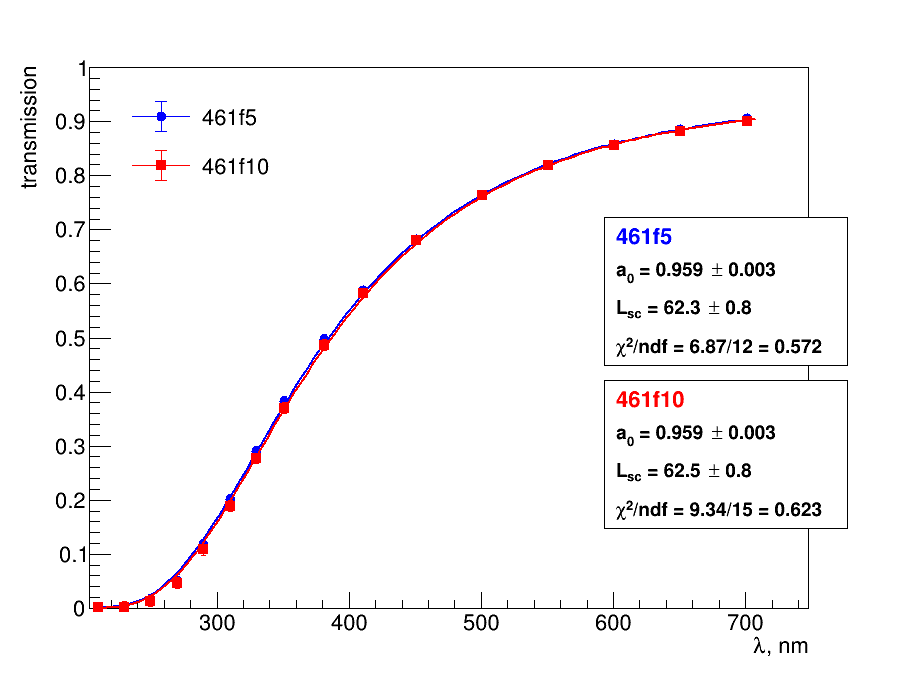}
\caption{The transmittance of the focusing aerogel tiles 461f5 (blue circles) and 461f10 (red squares). The thickness is 35~mm.}
\label{fig:fig8}
\end{figure}

Given that the Rayleigh scattering length in aerogels is significantly shorter than the absorption length, this optical parameter can be derived from transmittance measurements. The procedure of light scattering length measurement was described earlier~\cite{Buzykaev1999}. The scattering length \(L_{sc}\) of aerogel is determined from the wavelength-dependent light transmission, which is fitted by the Hunt formula:

\begin{equation}
T = \frac{I}{I_0} = a_0 \cdot \exp\left(-\frac{d}{L_{sc} \cdot (\lambda / 400)^4}\right),
\end{equation}

where \(T\) = \(I/I_0\) is the transmission, \(a_0\) accounts for surface losses, \(d\) is the sample thickness, \(\lambda\) is the wavelength, and \(L_{sc}\) is the scattering length at 400~nm. The \(\lambda^{-4}\) dependence reflects Rayleigh scattering in the aerogel structure. Examples of the measured transmittance for the four-layer aerogels 461f10 and 461f5 at a single point are shown in~Fig.~\ref{fig:fig8}. Table~\ref{tab:1} lists the weighted average light scattering lengths $L_{sc}$ at $\lambda = 400$~nm.

In aerogel RICH detectors, the Cherenkov angle resolution is affected by the surface planarity ($\Delta$) of the tiles, which is defined as the peak-to-valley surface variation normalized by the lateral dimension~\cite{Contalbrigo2017}. The planarity of the aerogel tiles was measured using a mechanical setup. The tile was placed on a measurement stand, and the height was recorded at multiple points across its surface according to a predefined grid. An alignment procedure was applied to correct for the global tilt of the tile by fitting and subtracting linear components along two orthogonal directions. After this correction, the planarity was defined as the difference between the maximum and minimum measured heights~\cite{Barnyakov2020}.

Novosibirsk-manufactured 30~mm-thick aerogel tiles exhibit exceptional surface planarity ($\Delta \leq 1\%$ of the lateral dimensions; for example, in the case of the tile measuring 200~$\times$~200~mm$^2$, the permissible planarity is $\leq$ 2~mm), meeting stringent detector specifications~\cite{Contalbrigo2017, Barnyakov2020}. The parameter variations within the small batch do not exceed the measurement uncertainties of a single sample.

\begin{table}[htb]
\centering
\caption{Parameters of four-layer monolithic aerogel manufactured in 2023: layer thickness $d_i$ (mm), refractive index $n_i$, mean light scattering length $L_{sc}$ (mm), and planarity $\Delta$ (mm).}
\label{tab:1}
\begin{tabular}{|c|c|c|c|c|}
\hline
	\textbf{aerogel} & \textbf{$d_i$} & \textbf{$n_i$} & \textbf{$L_{sc}$} & \textbf{$\Delta$} \\
\hline
461f5   & 9.0 & 1.045 $\pm$ 0.001 & 62.3 $\pm$ 0.3 & 1.32\\
        & 9.0 & 1.043 $\pm$ 0.002 &                &      \\
        & 8.5 & 1.040 $\pm$ 0.002 &                &       \\
        & 7.9 & 1.038 $\pm$ 0.001 &                &        \\
\hline
461f10  & 8.5 & 1.046 $\pm$ 0.001 & 62.5 $\pm$ 0.3 & 1.40\\
        & 8.5 & 1.043 $\pm$ 0.002 &                &      \\
        & 9.0 & 1.040 $\pm$ 0.002 &                &       \\
        & 9.0 & 1.039 $\pm$ 0.001 &                &        \\
\hline
\end{tabular}
\end{table}

These aerogel blocks were tested in detail using relativistic electrons at the BINP beam test facility~\cite{Abramov2016}. The single-photon resolution in the Cherenkov angle is about 7~mrad for a \text{3 $\times$ 3~mm$^2$} pixel in the photon detector. In this experiment, the pixel size was varied using a shading mask as described in~\cite{Barnyakov2020a}. After subtracting the pixel contribution in quadrature, the intrinsic resolution of the focusing aerogel tile is estimated to be 5.6~mrad. This value is mainly determined by the layer thicknesses (8–9~mm, contributing ~4~mrad) and refractive index variation (~4~mrad evaluated from GEANT4 simulations). The tests are test scheme and results presented in more detail in~\cite{Barnyakov2024}.

%%=============================================================================
\section{Thick aerogel radiator} 
\label{sec:thick}

Previously, the largest aerogel tiles from Novosibirsk were:
\begin{itemize}
    \item 200~$\times$~200~$\times$~30~mm$^3$ for aerogel with refractive index of 1.05~\cite{Barnyakov2020};
    \item 200~$\times$~200~$\times$~50~mm$^3$ for aerogel with refractive index of 1.03~\cite{Bellunato2009}.
\end{itemize}

\begin{figure}[htb]
\centering
\includegraphics[width=\linewidth]{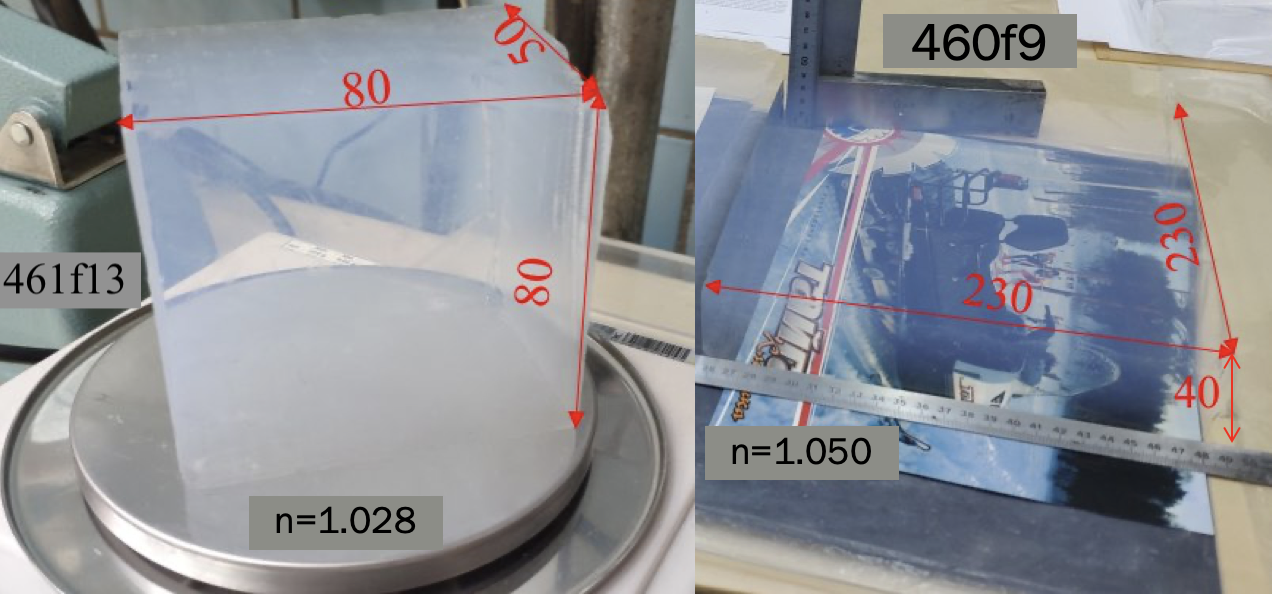}
\caption{Thick aerogel tiles:
  sample 461f13, the quarter-tile ($n = 1.028$, thickness 50~mm), on the left and 
 460f9 ($n = 1.050$, thickness 40~mm) on the right.
    }
\label{fig:fig4}
\end{figure}

\begin{figure}[htb]
\centering
\includegraphics[width=\linewidth]{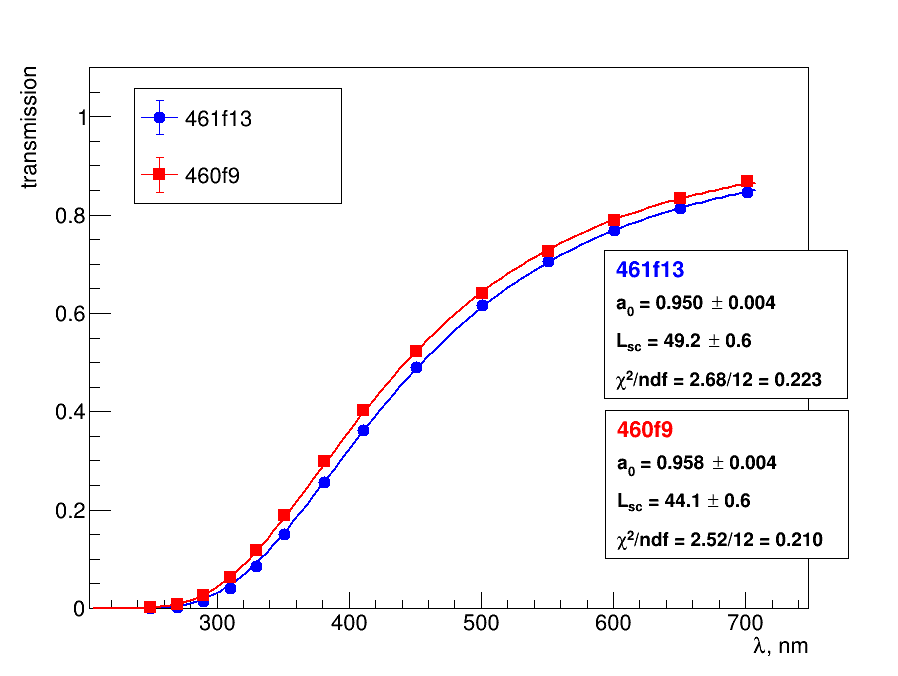}
\caption{
Transmittance of thick aerogel tiles: 
    461f13 ($n = 1.028$, thickness 50~mm, blue circles) and 
    460f9 ($n = 1.050$, thickness 40~mm, red squares).
}
\label{fig:fig9}
\end{figure}

Modifications to the primary annealing procedure allowed the production of aerogels with $n = 1.05$ of unprecedented thickness. Fig.~\ref{fig:fig4} (right) shows aerogel with a refractive index of 1.05 and a thickness of 40~mm — the record for this refractive index. Aerogel with a refractive index of 1.03 and a thickness of 50~mm was also reproduced, as shown in Fig.~\ref{fig:fig4} (left). Table~\ref{tab:2} summarises the parameters of aerogel tiles with lateral dimensions of 200~$\times$~200~mm. Examples of the measured transmittance for the thick aerogels 461f13 ($n = 1.028$, thickness 50~mm) and 460f9 ($n = 1.050$, thickness 40~mm) at a single point are shown in~Fig.~\ref{fig:fig9}.

\begin{table}[htb]
\centering
%\scriptsize
\caption{Parameters of thick aerogel tiles produced in 2022–2023: thickness $d$ (mm), refractive index $n$, weighted average light scattering length $L_{sc}$ (mm), and planarity $\Delta$ (mm).}
\label{tab:2}
\begin{tabular}{|c|c|c|c|c|}
\hline
	\textbf{aerogel} & \textbf{$d$} & \textbf{$n$} & \textbf{$L_{sc}$} & \textbf{$\Delta$} \\
\hline
461f4  & 40 & 1.046 & 44.5 $\pm$ 0.2 & 1.99\\
460f9  & 40 & 1.050 & 44.8 $\pm$ 0.3 & --\\
\hline
460f11 & 40 & 1.027 & 47.0 $\pm$ 0.4 & 1.90\\
461f14 & 40 & 1.028 & 50.1 $\pm$ 0.3 & --\\
\hline
460f15 & 50 & 1.027 & 43.6 $\pm$ 0.3 & --\\
461f13 & 50 & 1.028 & 50.0 $\pm$ 0.3 & 1.02\\
\hline
\end{tabular}
\end{table}

%%=============================================================================
\section{Conclusion}

Significant progress was achieved in 2022--2023 in the production of aerogel-based Cherenkov radiators for the RICH detectors. Based on the TGA/DSC analysis, the primary annealing procedure was optimized to minimize crack formation in large tiles. Monolithic four-layer focusing tiles (230~$\times$~230~$\times$~35~mm$^3$) with the target refractive indices and high transparency were produced for the first time.
The results show that the variation of these parameters within a small batch does not exceed the measurement uncertainty of a single sample.
In addition, large, highly transparent aerogel blocks with a refractive index of 1.05 and a thickness of 40~mm were produced, representing another achievement.

%%=============================================================================
\section*{Funding}
The works were supported in part by the project Mirror Laboratories of HSE University.

%%=============================================================================
\section*{Declaration of competing interest}
The authors declare that they have no known competing financial interests or personal relationships that could have appeared to influence the work reported in this paper.

%%=============================================================================
\section*{CRediT authorship contribution statement}
\textbf{A.Yu. Barnyakov:} Conceptualization, Investigation, Writing – review \& editing. \textbf{A.F. Danilyuk:} Methodology, Resources. \textbf{A.A. Kattsin:} Investigation, Validation, Data curation, Writing – original draft, Visualization. \textbf{E.A. Kravchenko:} Methodology. \textbf{A.A. Pochtare:} Investigation. \textbf{A.Yu. Predein:} Resources. 

%%=============================================================================
\bibliographystyle{elsarticle-num}
\bibliography{references}

\end{document}